# Designing Rating Systems to Promote Mutual Security for Interconnected Networks

Jie Xu, Yu Zhang and Mihaela van der Schaar

***Abstract*—Interconnected autonomous systems (ASs) often share security risks. However, an AS lacks the incentive to make (sufficient) security investments if the cost exceeds its own benefit even though doing that would be socially beneficial. In this paper, we develop a systematic and rigorous framework for analyzing and significantly improving the mutual security of a collection of ASs that interact frequently over a long period of time. Using this framework, we show that simple incentive schemes based on rating systems can be designed to encourage the ASs' security investments, thereby significantly improving their mutual security. When designing the optimal rating systems, we explicitly consider that monitoring the ASs' investment actions is imperfect and that the cyber-environment exhibits unique characteristics. An important consideration in this design is the heterogeneity of ASs in terms of both generated traffic and underlying connectivity. Our analysis shows that the optimal strategy recommended to the ASs on whether to make or not security investments emerges as a tradeoff between the performance gains achieved by ensuring the AS's compliance with the recommended strategy and the efficiency loss induced by the imperfect monitoring. When the monitoring errors are sufficiently small or the traffic and connectivity structure of the AS collection exhibits the "Maximal Critical Traffic (MCT)" property (i.e. the critical traffic of the whole collection is no less than that of any subset of the AS collection), it is optimal to recommend all ASs to make security investments. However, when this network property is not satisfied, an improved performance can be achieved when some ASs are recommended to NOT make security investments. Many simple network topologies (e.g. the complete, the "line", the "star" graphs and etc.) exhibit the "MCT" property. However, a common topology on the Internet - the "core-periphery" topology - does not possess the "MCT" property and in this case, we prove that whether or not it is optimal to recommend all ASs to make security investments depends on the AS collection size. Even though this paper considers a simplified model of the interconnected ASs' security, our analysis provides important and useful insights for designing rating systems that can significantly improve the mutual security of real networks in a variety of practical scenarios.**

## I. Introduction

A key focus of network science is ensuring the "health" of interconnected, interdependent complex networks [1][2][3]. For instance, autonomous systems (ASs) on the Internet are interconnected with each other and share security risks. Recent research notes that self-interested ASs are often tempted to reduce their own security investments in order to reduce their costs, i.e. they "free-ride" [1][4], due to the positive externalities of security investments (i.e. the security investment of an individual AS does not only enhance its own security, but also enhances the security of other ASs that connect with it). Therefore, a key challenge in maintaining the "health" of the Internet is how to design efficient incentive schemes to encourage cybersecurity investments from ASs. However, the extent to which these incentives can be and *should* be provided heavily depends on the traffic and connectivity structure of the AS collection as well as

The authors are with the Dept. of Electrical Engineering, University of California Los Angeles (UCLA), Los Angeles, CA, 90095, USA, E-mail: jiexu@ucla.edu, yuzhang@ucla.edu, mihaela@ee.ucla.edu.

the accuracy of monitoring ASs' investment actions.

We consider a collection of interconnected ASs that send traffic to each other across an underlying topology and aim at minimizing the overall security cost[2] of this AS collection. Each AS communicates with others. This communication is valuable but also dangerous, because it may contain malware. The individual AS can protect themselves by filtering inbound traffic. We want to set up a system in which AS's also protect each other by filtering outbound traffic. The problem is that filtering outbound traffic is costly for the sending AS and provides no direct benefit to the sending AS. To deal with this problem, we exploit the ongoing nature of the ASs' interaction to construct policies in which the current interactions depend on the past history of interactions within the collection and, in particular, on the extent to which the past behavior of an AS has been in accordance with the recommended behavior (e.g. make security investments and exchange secure traffic). For this, we build on the general theory of repeated games, but with many necessary innovations (discussed below). Our approach is based on *rating systems*. An AS is rated according to its past security investment actions. The other ASs reward/punish it by employing different security policies (which result in different outbound traffic filtering qualities) for the traffic sent to this AS depending on its rating. The proposed rating system consists of four components: (1) The rating set is taken to be a finite (totally/partially) ordered set. (2) The monitoring component makes observations and receives reports from each AS about that AS's experiences, including traffic analytics, performance, infection ratios, etc. (3) On the basis of this information and the current ratings, the assessment component updates the rating of each AS. (4) The recommendation component reports the ratings to the ASs together with recommended strategies. These recommendations include how each AS should install/update its own security policies (e.g. invest in certain technologies or not) and how it should interact with other ASs by allowing, denying or limiting service depending on their ratings.

The rating system itself is implemented by a Rating Agency (RA), which might be operated by a private entity or a governmental agency. In implementing the rating system, the RA carries out (some) direct observations, collects reports, aggregates these observations and reports into ratings and uses these ratings to recommend security policies for the ASs to follow. Note that the

[2] A rigorous definition of security costs will be provided in Section II.

RA itself typically has no power to enforce these recommended policies – they are merely suggestions to the ASs – so implementation of policies is completely decentralized. The ASs will actually execute the recommended policies only if the rating system is designed in such a way that it provides the incentives to the ASs that lead them to prefer compliance rather than deviation. We say a rating system is *incentive-compatible (IC)* if it has this property.

**Imperfect Monitoring**: In practice, monitoring of ASs' investment actions is never perfect and hence, the rating assessment of ASs is not always accurate. This makes the design problem much more complicated. (1) If monitoring and security were perfect, the RA could employ the strongest punishment (e.g. a trigger strategy[3] [21]) against deviation and hence, it would provide the strongest incentive for ASs to comply rather than deviate. However, when monitoring is imperfect, reports cannot be guaranteed to be accurate and compliant behavior cannot be guaranteed to yield secure results. Relying on a strong punishment would therefore lead to very low security if trigger strategies were employed. Instead, choosing milder punishments when the deviation of an AS is reported becomes more efficient. However, this punishment should not be too mild such that sufficient incentives are provided to the ASs to comply rather than deviate. Therefore, the optimal punishment needs to be designed as a tradeoff between efficiency and incentive-compatibility. (2) Even if the monitoring technology is fixed exogenously, the *level* of monitoring will typically be a design parameter. One important design parameter is the time length (which is determined by the rating assessment period) during which the monitoring is performed. Too frequent updates not only lead to too many messaging and computational overheads but also reduce the monitoring accuracy. Typically, a longer assessment period results in more accurate monitoring reports, but too long an assessment period makes the punishment come too late to be effective due to ASs' discounting of future utilities. Therefore, the optimal rating assessment frequency is a tradeoff between the monitoring accuracy and ASs' valuation of future utilities.

**Traffic and Connectivity Heterogeneity**: An important distinct characteristic of the cyber-environment is that ASs are heterogeneous in terms of generated traffic and underlying connectivity. A certain punishment chosen by the RA may provide strong enough incentives for some ASs to comply but may fail (i.e. may lead to deviations) by other ASs. As we emphasized before, when the monitoring is imperfect, a strong punishment induces a significant efficiency loss. Hence, if a strong punishment is required for all ASs to follow the recommended strategy (e.g. make security investment), then the performance of the collection of ASs may be improved if the RA can use a milder punishment by recommending some ASs to *not* make investments. Therefore, the optimal recommended strategy is selected as a tradeoff between the performance gains achieved by ensuring the AS's compliance with the recommended strategy and the efficiency loss induced by the imperfect monitoring. Our analysis shows that when monitoring errors are small and when the traffic and connectivity of the AS collection exhibits the "Maximal Critical Traffic" property, the optimal strategy is to recommend all ASs to make security investments. However, when the network environment does not satisfy these conditions, an improved mutual security may be achieved when some ASs are recommended to *not* make security investment.

[3] For instance, following any point in time at which there is any evidence that any AS had deviated from any previous recommendation, the RA recommends that each AS in the collection do not deploy OTC afterwards.

The rest of this paper is organized as follows. Section II introduces the repeated security investment game and formulates the rating system design problem. Section III provides the optimal design for a given recommended strategy. Section IV determines the optimality conditions of the recommended strategies. An efficient algorithm to compute the optimal design is also proposed. Section V provides illustrative results to highlight the features of the proposed rating system. Section VI discusses related works. Section VI concludes the paper.

## II. System Model

### *A. The security investment game*

We consider a collection of $N > 1$ inter-connected autonomous systems (ASs) [4], represented by a set $\mathcal{N} = \{1,2,...N\}$. The ASs send traffic to each other. Let $\Lambda$ be the $N \times N$ traffic matrix, with $\lambda_{i,j} \geq 0$ being the traffic rate sent from AS $i$ to AS $j$. And we assume $\lambda_{i,i} = 0$, $\forall i \in \mathcal{N}$. The aggregated outbound/inbound traffic of AS $i$ are $\mu_i = \sum_{k=1}^{N} \lambda_{i,k}$ and $\nu_i = \sum_{k=1}^{N} \lambda_{k,i}$, respectively.

Traffic contains malware (e.g. spam, viruses etc.) and hence, the ASs deploy traffic control technology (e.g. firewalls, traffic filters) to protect their own devices. In a passive protection system, the ASs only deploy inbound traffic control. In a proactive protection system, the ASs also deploy *outbound traffic control* (OTC) to enhance the cybersecurity [7][9]. Such control can involve, for example, outbound filters which can block malware as it leaves the AS. OTC is much more effective because the ASs have improved control over their own hosts and devices and the traffic originating from them [7][9]. Fig. 1 illustrates a system with OTC deployment.

[4] We do not consider the traffic routing problem and just focus on the direct traffic exchanged among adjacent ASs (i.e. one-hop traffic exchanges), but our work can be generalized to assist multi-hop routing etc.

We assume that without the deployment of OTC, each unit of the traffic contains malware (e.g. spam, viruses etc.) with a probability $\overline{p} \in [0,1]$. On the other hand, OTC can reduce this probability to some lower value $p \in [\underline{p}, \overline{p}]$ that is determined by the specific OTC policy (e.g. by employing a protocol-level malware scanner [8]). We assume that $\underline{p}$ is the lowest probability that OTC is able to achieve due to technology constraints. Network traffic filters provide protection by sampling packets or sessions and either comparing their contents to known malware signatures or looking for anomalies likely to be malware. By adjusting different sampling rates, the OTC technology is able to achieve different filtering qualities (i.e. different probabilities that the traffic contains malware) [10]. When the traffic contains malware, the receiving AS has to take recovery measures (e.g. cleaning the traffic, patching or replacing with new devices), with $c_r$ being the average recovery cost per unit traffic that contains malware.

In the security investment game, time is divided into periods of length $T$. At the beginning of each period, each AS is strategic in choosing to deploy OTC or not by taking an action $a \in \mathcal{A} = \{0,1\}$, where “1” stands for “invest” (deploy OTC) and “0” stands for “not invest” (not deploy OTC). The cost of deploying OTC is $c_l$ per unit time (as in [10]). Therefore, the cost per period of deploying OTC is $c_l T$. Without loss of generality, we normalize the costs and let $c_r = 1$, $c_l = c$. Nevertheless, our framework can also be easily extended to study different cost types. For example, the deployment cost may consist of an initial implementation cost and incremental costs which depend on the sending traffic amount.

ASs are long-lived and discount the future utility at a constant rate. Specifically, the utility of the next period is discounted as in [20] by $e^{-\beta T}$ where $\beta \in \mathbb{R}^+$ is the discount factor parameter. We assume that the utility within a period is not discounted but our analysis also applies to this discounting case because the ratio of the recovery cost to the deployment cost remains unchanged. Within a period, the deployment of OTC is not in an AS's self-interest since it entails investment expenditures but does not result in immediate benefits (enhancement on the security level) for this AS. Therefore, the dominant strategy of ASs is $a = 0$ for each period without proper incentives being provided.

*B. The rating system*

Rating systems are designed to provide the ASs with OTC deployment incentives. Before the rating system is implemented, the RA designs a rating protocol $\kappa$ which consists of four components: a set of ratings $\Theta$, a recommended strategy $\sigma$, a monitoring technology $\chi$ and a rating assessment rule $\phi$, aiming to minimize the overall security costs (including investment costs and recovery costs) of the AS collection. Fig. 2 illustrates the interactions between four ASs in a rating system. The rating system has the following components.

**(1) Rating Set.** The rating set that we consider in this paper is a binary set $\Theta = \{0,1\}$. However, our analysis can be easily extended to multiple rating levels[5]. Each AS is assigned with a rating label according to its past security investment actions.

**(2) Recommended strategy.** The RA recommends different OTC investment strategies (for brevity, we simply call this the *recommended strategy*) for the ASs. The recommended strategy is a mapping $\sigma: \mathcal{N} \to \mathcal{A}^N$ where $\sigma(\mathcal{N})$ is the recommended action profile (i.e. a vector) for all ASs. Because $\mathcal{A}$ is a binary space (i.e. deploy or not), the recommended strategy recommends some ASs to deploy OTC and recommends the rest of them not to deploy OTC. Denote $\mathcal{P}$ as the space of all subsets of the AS collection $\mathcal{N}$. Then for each $\sigma$, it is associated with a subset $P \in \mathcal{P}$ of the AS collection as follows,

$$a_i = \begin{cases} 1, & \text{if } i \in P \\ 0, & \text{if } i \in \mathcal{N} \setminus P \end{cases}$$

Therefore, we equivalently write this strategy as $\sigma_P$. We call $\sigma_{\mathcal{N}}$ the *full deployment strategy* (FDS, i.e. the strategy that recommends all ASs to deploy OTC) and all the other strategies are referred to as *partial deployment strategies* (PDSs). Importantly, note that these strategies are not binding contracts that the ASs must follow but only recommendations. Hence, the RA needs to design an IC rating system which makes it in the self-interest of the ASs to follow these recommendations.

[5] In some deployment scenarios, we can even prove that the binary set is in fact the optimal set.

Once the OTC is deployed by an AS, this AS can choose different filtering qualities (e.g. using different traffic sampling rates), which are measured by the resulting probability $p \in [\underline{p}, \overline{p}]$ that the traffic contains malware for the *receiving* ASs to which it sends traffic. The RA also recommends a set of filtering qualities that depend on only the receiving ASs' ratings[6]. Since the rating set is binary, these recommendations are $p_1, p_0$ for the receiving ASs of $\theta = 1$ and $\theta = 0$, respectively. We assume $\underline{p} \le p_1 \le p_0 \le \overline{p}$ because the traffic of the receiving ASs of high ratings should receive better quality. Because the OTC deployment cost does not depend on the specific set of filtering qualities, the ASs would just (weakly) follow any filtering qualities recommended by the RA and so $p_1, p_0$ are design parameters of the RA [7]. However, the ASs can choose whether or not to deploy OTC. If the AS does not deploy OTC (either because it deviates from a recommendation of deploying or it follows a recommendation of not deploying), the filtering quality for all its outbound traffic is $\overline{p}$. One might think that the RA should simply choose $p_0 = \overline{p}$. However, as we will show later, strong punishments may induce significant efficiency loss when the monitoring is imperfect. Therefore, designing the filtering qualities (e.g. the traffic sampling rates) is critical for the efficiency of the rating system.

**(3) Monitoring technology**. Let $\mathcal{S} = \{0,1\}$ be the set of signals observable by the RA about an individual AS. A monitoring technology is a measure-valued map $\chi: \mathcal{A} \times \mathcal{A} \to \Delta(\mathcal{S})$, where $\chi(s \mid a, \tilde{a})$ denotes the conditional probability that $s$ was observed given that $a$ was played by an AS when the recommended action is $\tilde{a}$. Define $\epsilon_{a,\tilde{a}}(T)$ as the monitoring error probability and,

$$\chi(0 \mid 1,1) = \epsilon_{1,1}(T), \quad \chi(0 \mid 0,0) = \epsilon_{0,0}(T), \quad \chi(1 \mid 1,0) = \epsilon_{1,0}(T), \quad \chi(1 \mid 0,1) = \epsilon_{0,1}(T), \tag{1}$$

and $\chi(0 \mid a, \tilde{a}) = 1 - \chi(1 \mid a, \tilde{a})$. To simplify the notations, in this paper we assume the monitoring technology is identical for all ASs and has symmetric error patterns $\epsilon_{a,\tilde{a}}(T) = \epsilon(T), \forall a, \tilde{a}$. However, our analysis can be easily extended to the asymmetric cases. Therefore, the monitoring technology accurately detects compliance ($s = 1$) or deviation ($s = 0$) with probability $1 - \epsilon(T)$. We make the following assumption on $\epsilon(T)$.

[6] The RA could also recommend specific filtering qualities for specific receiving ASs. This may improve the system performance but significantly increases the rating system complexity. In this paper, we focus on only rating-dependent recommendations and show that even this simple method can significantly enhance the mutual security.

[7] If costs also depend on filtering qualities, then the RA needs to consider additional incentive constraints. This is a much more complex problem and requires future research effort. Alternatively, we assume that the technology is given (selected) by the RA and that the ASs cannot manipulate the specific parameters.

**Assumption 1**: $\epsilon'(T) \leq 0, \epsilon''(T) \geq 0, \lim_{T \to \infty} \epsilon(T) = 0, \epsilon(0) \leq 0.5$.

Assumption 1 states that the monitoring error decreases with the period since an increased number of observations can infer the real action with a higher accuracy. As this period length goes to infinity, the monitoring becomes perfectly accurate. However, the marginal improvement in terms of monitoring accuracy decreases as the period length is increased. Moreover, we also require that the largest monitoring error is no larger than 0.5 such that monitoring is at least useful.

**(4) Rating assessment rule**. The rating assessment rule, which is performed at the end of each period, decides how the ratings of the ASs should be updated according to the monitored signal, $\phi : \Theta \times \mathcal{S} \to \Theta$, where $\phi(\theta \mid s)$ is the next period rating conditional on the observed signal $s$ if the current rating is $\theta$. Particularly, we consider the following assessment rules.

$$\phi(\theta \mid s = 1) = 1, \qquad \phi(\theta \mid s = 0) = 0 \tag{2}$$

Therefore, an AS that complies with the recommended strategy receives a high rating while an AS that deviates from the recommended strategy receives a low rating. Note that even if an AS does not deploy OTC, its rating still can be high if the recommended strategy for it is "not deploy" because this AS *is complying* with the recommended strategy. We could just assign the ASs outside $P$ with fixed high ratings regardless of their signals, but using (2) has advantages in terms of analysis simplicity and does not change our results.

*C. The design problem*

The objective of the RA is to minimize the overall security costs of the ASs in the collection per unit time. The security cost includes the recovery cost and the investment cost. For an AS $i$, its security cost is $J(i \mid \kappa) = p(i \mid \kappa)\nu_i + [\sigma]_i c$, where $p(i \mid \kappa)$ is the average probability that the inbound traffic of AS $i$ contains malware when the rating system $\kappa$ is used and is IC. The overall security cost of the AS collection is thus: $J(\kappa) = \sum_{i \in \mathcal{N}} J(i \mid \kappa)$. The RA selects the design parameters, i.e. the recommended strategy $\sigma$, the filtering qualities $p_0, p_1$ and the rating assessment period $T$, according to the environment parameters $\mathcal{N}, c$ and the traffic matrix $\Lambda$ to minimize the overall security cost:

$$\text{(DESIGN)} \qquad \begin{array}{ll} \underset{\kappa = \{T, p_0, p_1, \sigma\}}{\text{minimize}} & J(\kappa) \\ \text{subject to} & \text{the rating system } \kappa \text{ is IC} \end{array}$$

We make the following assumption on the AS collection.

**Assumption 2**: $c < (\overline{p} - \underline{p})\max\{\nu_i,\mu_i\},\forall i \in \mathcal{N}$ .

Assumption 2 indicates that: (1) The overall security cost is minimized when all ASs deploy OTC; (2) For each AS, its investment cost is no larger than the maximum benefit that it can possibly obtain from OTC and hence, the rating system may be able to provide incentives for all ASs to deploy OTC. Note that if Assumption 2 does not hold for some ASs in a collection, then it is optimal for the RA to recommend them not to deploy OTC and these ASs would just follow the recommendation. Hence, these are only trivial cases for our study and we will eliminate them from our analysis. Under Assumption 2, the "first-best" performance (i.e. the minimal overall security cost per unit time), denoted by $J^{**}$ , is achieved when all ASs *cooperatively* invest in OTC. Therefore,

$$J^{**}=\underline{p}\sum_{i\in\mathcal{N}}\mu_i + Nc$$

In the remaining part of this paper, we design rating systems that aim at minimizing $J(\kappa)$ for *self-interested* ASs by recommending OTC deployment strategies. We consider the following three benchmarks (designs of the AS collection) which will be evaluated in the simulations:

1. There is no OTC deployed in the AS collection ("no OTC").
2. The filtering quality of the OTC is not rating-based ("rating-independent OTC").
3. Receiving ASs with low (high) ratings receive the worst (best) quality $\overline{p}$ ( $\underline{p}$ ) ("worst-best").

We briefly discuss the design flow of the rating system. For notational simplicity, we abuse the notation and denote $J(\sigma_P)$ as the overall security cost for a fixed recommended strategy $\sigma_P$ .

1. We fix a recommended strategy $\sigma_P$ and optimize the other design parameters $p_0^*, p_1^*, T^*$ such that the overall security cost is minimized for $\sigma_P$ , denote by $J^*(\sigma_P)$ (in Section III).
2. We determine the optimality conditions for the recommended strategies (in Section IV). Particularly, we determine the structural properties when FDS is optimal. In general scenarios, an efficient algorithm to compute the optimal strategy is proposed to reduce the problem complexity from $O(2^N)$ to $O(N)$ .

## III. Optimal Design for a Fixed Strategy

In this section, we study the optimal rating system design by fixing a recommended strategy

$\sigma_P$. Therefore, the only design parameters are $p_0, p_1, T$ and the objective function is the overall security cost $J(\sigma_P)$. In the following, we define the *critical traffic* of a subset $P$ of the AS collection, which will play a critical role for the rating system design problem.

**Definition 1**: Consider any subset $P$ of the AS collection $\mathcal{N}$. For any AS $i \in P$, let $\nu_i(P)$ be its aggregate inbound traffic originating from all ASs *in* $P$. The critical traffic of $P$ is $\underline{\nu}(P) = \min_{i \in \mathcal{P}} \nu_i(P)$. (Note that $\mathcal{N}$ is also a subset of $\mathcal{N}$).

Note that when calculating the critical traffic of a subset $P$, we only consider the traffic between the ASs within $P$ but not that outside $P$.

*A. Individual AS's IC constraints*

The long-term utility of an AS $i$ of rating $\theta$ is defined as the discounted sum of its current utility and expected future utility. If all the ASs comply to the recommended strategy $\sigma_P$ and $\sigma_P(\mathcal{N}) = \mathbf{a}_P$, the long-term utility can be expressed as $\forall \theta \in \Theta$,

$$U_i^\infty(\theta \mid \mathbf{a}_P) = -(p_\theta \nu_i^P + \overline{p}(\nu_i - \nu_i^P) + c a_{P,i})T \\ - e^{-\beta T}[(1-\epsilon)U_i^\infty(\phi(\theta \mid 1) \mid \mathbf{a}_P) + \epsilon U_i^\infty(\phi(\theta \mid 0) \mid \mathbf{a}_P)]$$

The first part of this utility $(p_\theta \nu_i^P + \overline{p}(\nu_i - \nu_i^P) + c a_{P,i})T$ is the security cost incurred in the current period if AS $i$'s rating is $\theta$ and the second part $e^{-\beta T}[(1-\epsilon)U_i^\infty(\phi(\theta \mid 1) \mid \mathbf{a}_P) + \epsilon U_i^\infty(\phi(\theta \mid 0) \mid \mathbf{a}_P)]$ is the discounted utility (cost) in the subsequent periods.

Alternatively, if AS $i$ unilaterally deviates from $\sigma_P$ by choosing $\tilde{a}_i \neq a_{P,i}$, its long-term utility becomes,

$$U_i^\infty(\theta \mid \tilde{a}_i, \mathbf{a}_{P,-i}) = -(p_\theta \nu_i(P) + \overline{p}(\nu_i - \nu_i(P)) + c\tilde{a}_i)T \\ - e^{-\beta T}[\epsilon U_i^\infty(\phi(\theta \mid 1) \mid \mathbf{a}_P) + (1-\epsilon)U_i^\infty(\phi(\theta \mid 0)) \mid \mathbf{a}_P]$$

**Proposition 1**: The rating system with $\sigma_P$ is IC for AS $i$ if and only if $\forall \theta \in \Theta$,

$$(1-2\epsilon)e^{-\beta T}[U_i^\infty(\phi(\theta \mid 1) \mid \mathbf{a}_P) - U_i^\infty(\phi(\theta \mid 0) \mid \mathbf{a}_P))] \geq (2a_{P,i} - 1)cT \quad (3)$$

**Proof**: If the rating system is IC, then clearly there are no profitable one-shot deviations. We can prove the converse by showing that if the rating system is not IC, there is at least one profitable one-shot deviation. Since $cT$ is bounded, this is true by the "one-shot" deviation principle in the repeated game theory [21]. □

Proposition 1 is based on the "one-shot" deviation principle [21] and shows that if an AS cannot gain by unilaterally deviating from $\sigma_P$ only in the current period and following $\sigma_P$ afterwards, it cannot gain by switching to any other strategies (possibly multiple-shot deviations) either, and vice versa. The right-hand side (RHS) of (3) is the current gain by deviating, while the left-hand side (LHS) of (3) represents the discounted expected future loss due to such deviations. Hence, an AS has no incentive to deviate if and only if its future loss outweighs its current gain upon deviation.

For the ASs which are not in $P$, it is easy to see that the rating system is always IC because LHS of (3) is positive while the RHS of (3) is negative. Hence, these ASs will just follow the recommended strategy and do not deploy OTC. However, whether the ASs in $P$ will follow the recommended strategy depends on the rating system parameters which we discuss next.

*B. Incentive-compatible (IC) region*

Before we design the optimal rating system parameters, it is important to know whether there exist rating systems such that a recommended strategy $\sigma_P$ can be IC. In particular, we are interested in characterizing the region of the discount factor parameter $\beta$ that such IC rating system that recommends $\sigma_P$ exist, denoted by $\Pi(\sigma_P)$. Theorem 1 provides a condition when $\Pi(\sigma_P)$ is the largest.

**Theorem 1**: If $\epsilon(0) \leq \frac{1}{2}\left(1 - \frac{c}{(\overline{p} - \underline{p})\underline{\nu}(P)}\right)$, then $\Pi(\sigma_P) = \mathbb{R}^+ \setminus \infty$.

**Proof**: According to (3), a rating system is IC if and only if

$$(1 - 2\epsilon(T))e^{-\beta T}(p_0 - p_1)\underline{\nu}(P)T \geq cT \tag{4}$$

Therefore, the discount parameter must satisfy,

$$\beta \leq \frac{1}{T}\ln\frac{(1 - 2\epsilon(T))(p_0 - p_1)\underline{\nu}(P)}{c} \tag{5}$$

If the monitoring error is small enough, then the term inside the $\ln$ function is larger than 1. Because $\epsilon(0), p_0, p_1, \underline{\nu}(P), c$ are bounded, for any $\beta \in \mathbb{R}^+ \setminus \infty$, we can always find $T$ small enough such that (5) holds. □

Theorem 1 states an important result for the existence of an IC rating system. If the monitoring technology is accurate enough, then for any environment setting (any discounting and any traffic matrix) there exists an IC rating system that recommends $\sigma_P$. However, given a recommended strategy and a monitoring technology, even though a rating system is IC, it can still induce significant efficiency loss without careful choices of the design parameters.

*C. Optimal design parameters*

Denote $\Omega(\beta,\sigma_P)$ as the set of IC rating systems given the discount factor parameter $\beta$ and the recommended strategy $\sigma_P$. If $\Omega(\beta,\sigma_P) \neq \varnothing$, then there exists at least one IC rating system. In this subsection, we determine the optimal design parameters of the rating systems that are IC. To do this, we first compute the security cost for a rating system $\kappa(T,p_0,p_1) \in \Omega(\beta,\sigma_P)$.

$$J(\sigma_P) = [(1-\epsilon(T))p_1 + \epsilon(T)p_0]\sum_{i\in P}\mu_i + \overline{p}\sum_{i\notin P}\mu_i + |P|c \qquad (6)$$

The first term is the recovery cost for the traffic originating from the set $P$; the second term is the recovery cost for the traffic originating from the remaining set; and the last term is the OTC deployment cost of the ASs in $P$. The optimal rating system minimizes (6) among all IC rating systems for given $\beta$ and $\sigma_P$.

**Theorem 2**: For all $\kappa(T,p_0,p_1) \in \Omega(\beta,\sigma_P)$, the optimal design parameters satisfy

1. $\max\{T:\kappa(T,p_0,p_1) \in \Omega(\beta,\sigma_P)\} < \frac{1}{\beta}\ln\frac{(\overline{p}-\underline{p})\underline{\nu}(P)}{c}$, and $T^* = \underset{T:\kappa(T,p_0,p_1)\in\Omega(\beta,\sigma_P)}{\arg\min}\; e^{\beta T}\frac{\epsilon(T)}{1-2\epsilon(T)}$
2. $p_1^* = \underline{p}$ , $p_0^* = e^{\beta T^*}\frac{c}{(1-2\epsilon(T^*))\underline{\nu}(P)} + \underline{p}$

**Proof**: By Proposition 1, if a rating system is IC, then

$$(1-2\epsilon(T))e^{-\beta T}(p_0 - p_1)\nu_i(P)T \geq cT, \forall i \in P \qquad (7)$$

Therefore,

$$e^{\beta T} \leq \frac{(1-2\epsilon(T))(p_0-p_1)\nu_i(P)}{c} < \frac{(\overline{p}-\underline{p})\nu_i(P)}{c}, \forall i \in P$$

Consider all $i \in P$, the upper bound of $T$ depends on $\underline{\nu}(P)$ and is given as in the first statement.

Given an rating assessment period $T$ such that $\kappa(T,p_0,p_1) \in \Omega(\beta,\sigma_P)$, we need to minimize both $p_1$ and $p_0$ in order to minimize (6). By (7),

$$p_0 \geq e^{\beta T}\frac{c}{(1-2\epsilon(T))\nu_i(P)} + p_1, \forall i \in P \qquad (8)$$

Therefore, for a fixed $T$, (6) is minimized when $p_1 = \underline{p}$ and $p_0$ is chosen as the binding value of (8) for the critical traffic. We then substitute these filtering qualities values into (6) and obtain

$$J(\sigma_P) = \left[\underline{p} + e^{\beta T}\frac{\epsilon(T)}{1-2\epsilon(T)}\frac{c}{\underline{\nu}(P)}\right]\sum_{i\in P}\mu_i + \overline{p}\sum_{i\notin P}\mu_i + |P|c \qquad (9)$$

To minimize (9), the optimal rating assessment period is therefore given by the first statement.

The optimal filtering qualities are therefore chosen for $T^*$ according to (8). □

Theorem 2 has several essential implications. The first statement indicates that increasing the rating assessment period has two opposite effects on the efficiency loss factor. On one hand, the RA wants to choose a longer $T$ because it reduces the monitoring error (i.e. $\epsilon(T)$ decreases with $T$) and hence, there is a smaller probability that an AS has a lower rating. On the other hand, if the rating assessment period $T$ is too long, then the punishment comes too late to be effective due to ASs' discounting of the future utilities (i.e. $e^{\beta T}$ increases with $T$). This requires a stronger punishment (higher $p_0^*$) and hence, it induces more efficiency loss for the low-rated ASs. Importantly, if $T$ is too long to provide strong enough punishment (due to the upper bound $\overline{p}$), then the rating system is not IC. Therefore, there is an upper bound on $T$ and the RA does not want $T$ to be very long. The optimal rating assessment period is thus a tradeoff between the monitoring capability and ASs' valuation of the future utility. We define

$$g(T)=e^{\beta T}\frac{\epsilon(T)}{1-2\epsilon(T)}$$

as the *efficiency loss factor* which determines the efficiency loss of a rating system due to monitoring errors. We denote $g_P^*(T)$ as the *minimum efficiency loss factor* given $\sigma_P$.

The second statement indicates the tradeoff of the punishment between incentive-compatibility and the efficiency loss. Low-rated ASs are punished for not following the recommended strategy. To make the rating system IC, the punishment should be strong enough such that the ASs in $P$ do not have incentives to deviate. Importantly, the incentives of the ASs in $P$ with the critical traffic $\underline{v}(P)$ are the most difficult to be satisfied since the relative benefits that they can receive from OTC are the smallest. Therefore, the optimal filtering qualities are the binding values of the IC constraints. This can be neither too small to satisfy the incentive constraints nor too large such that it induces additional efficiency loss.

## IV. Optimal Recommended Strategy

In the previous section, we determined the optimal rating system design given a recommended strategy $\sigma_P$. However, it was not clear which recommended strategy will lead to the minimal overall security cost among all possible strategies. In this section, we determine the optimal recommended strategy.

*A. Optimality of the strategy with full deployment*

The recommended strategy space is huge. For an AS collection with $N$ ASs, the cardinality of this space is $2^N$. Searching in this space is thus a demanding task for the RA. Intuitively, the optimal strategy seems to be $\sigma_{\mathcal{N}}$ (FDS), i.e. all ASs are recommended to deploy OTC, since the "first-best" is achieved when all ASs *cooperatively* deploy OTC. If this intuition would be correct, then the RA can simply choose $\sigma_{\mathcal{N}}$ to be the recommended strategy and design the corresponding optimal rating system and this can significantly reduce the complexity of solving the rating system design problem. Interestingly, we will show later that this intuition is not correct in certain network scenarios. Therefore, it is important to understand when $\sigma_{\mathcal{N}}$ is indeed the optimal strategy. The following proposition provides a sufficient condition with respect to the *minimum* efficiency loss factor $g^*_{\mathcal{N}}(T)$ for $\sigma_{\mathcal{N}}$ to be optimal.

**Theorem 3**: If the minimum efficiency loss factor is small enough, i.e.

$$g^*_{\mathcal{N}}(T) \leq \frac{(\overline{p} - \underline{p})\min_{i\in\mathcal{N}} \mu_i - c}{\sum_{i\in\mathcal{N}} \mu_i} \frac{\underline{\nu}(\mathcal{N})}{c}$$

then $\sigma_{\mathcal{N}}$ is the optimal recommended strategy.

**Proof**: Consider any PDS that recommends $M$ ASs not to deploy OTC. Their total outbound traffic is $\mu_M$. The maximal reduction of the recovery cost for the traffic originating from the ASs that are recommended to deploy OTC is

$$g^*_{\mathcal{N}}(T)\frac{c}{\underline{\nu}(P)}\left(\sum_{i\in\mathcal{N}} \mu_i - \mu_M\right) \tag{10}$$

The increased security cost for the other two parts in (6) is

$$\left(\overline{p} - \underline{p} - g^*_{\mathcal{N}}(T)\frac{c}{\underline{\nu}(P)}\right)\mu_M - Mc \tag{11}$$

The condition in the statement guarantees that (10) is smaller than (11). Therefore, the FDS must yield the minimal security cost. □

The efficiency loss occurs because (1) the RA has to choose a lower $p_0$ for low-rated ASs as a punishment to deter them from deviating from the recommended strategy and (2) the monitoring is imperfect and hence, this punishment is always carried out with a positive probability. If by recommending a PDS, the RA is able to choose a smaller $p_0$ but the rating system is still IC, then the recovery cost for the traffic originating from ASs in $P$ may be reduced. Moreover, by recommending some AS not to deploy OTC, the deployment cost is also reduced. On the other hand, a PDS recommends the ASs outside $P$ not to deploy OTC and their outbound traffic contains malware with a constant high probability which leads to a higher recovery cost for their receiving ASs. If the maximum reduction of the recovery cost for the traffic originating from ASs in $P$ plus the deployment cost is less than the minimum increased recovery cost for the traffic originating from ASs outside $P$, then changing to any PDS cannot reduce the overall security cost and hence, $\sigma_{\mathcal{N}}$ is optimal. (See Fig. 3)

Theorem 3 unravels the impact of the efficiency loss factor on the performance of the FDS. In fact, according to (9), the optimal performance of a rating system with the FDS can actually asymptotically achieve the "first-best" performance $J^{**}$ as $g^*_{\mathcal{N}}(T)$ goes to 0.

*B. "Maximal Critical Traffic" Property*

Theorem 3 implies that if the monitoring accuracy is high enough, then the FDS is optimal. Particularly, the FDS achieves the "first-best" if the monitoring is perfect. However, monitoring is never perfect in practice and therefore $g^*_{\mathcal{N}}(T)$ cannot be 0. In some scenarios, it could even be relatively large if the monitoring technology is not good enough. If the condition in Proposition 2 is not satisfied and the punishment is strong ($p^*_0$ is high), then the FDS may induce a significant efficiency loss. Therefore, it is possible that a PDS achieves a better performance. In this subsection, we determine a structural result of the AS collection such that the FDS is optimal.

**Assumption 3**: $g^*_{\mathcal{N}}(T) \le \min_{i\in\mathcal{N}} \left\{ \frac{(\overline{p} - \underline{p})u_i - c}{u_i} \frac{\underline{v}(\mathcal{N})}{c} \right\}$.

Assumption 3 indicates that by using the FDS $\sigma_{\mathcal{N}}$, the reduced recovery cost for the traffic originating from any individual AS is larger than the deployment cost of this AS. Therefore, the individual AS deploying OTC is socially beneficial. Note that the condition on the minimum efficiency loss factor in Assumption 3 is mild (Fig. 4 compares it with the bound established in Theorem 3) and hence, the minimum efficiency loss factor in fact can be quite large.

We first consider the minimum security costs of two different recommended strategies.

**Proposition 2**: Consider a strategy $\sigma_P$ such that $\underline{v}(P) \ge \underline{v}(\mathcal{N})$. Under Assumption 3, for any strategy $\sigma_{P'}$ such that $P' \subset P$, if $\underline{v}(P') \le \underline{v}(P)$, then $J^*(\sigma_{P'}) > J^*(\sigma_P)$.

**Proof**: Because $\underline{v}(P) \ge \underline{v}(\mathcal{N})$, it is easy to see that $g^*_P(T) \le g^*_{\mathcal{N}}(T)$. By Assumption 3, we obtain

$$g_P^*(T) \le \min_{i\in P}\left\{\frac{(\overline{p}-\underline{p})u_i - c}{u_i}\frac{\underline{\nu}(P)}{c}\right\} \quad (12)$$

By using $\sigma_{P'}$, the security cost difference is

$$J^*(\sigma_{P'}) - J^*(\sigma_P)$$
$$= \left[g_{P'}^*(T)\frac{c}{\underline{\nu}(P')} - g_P^*(T)\frac{c}{\underline{\nu}(P)}\right]\sum_{i\in P'}\mu_i + \left[\overline{p} - \underline{p} - g_P^*(T)\frac{c}{\underline{\nu}(P)}\right]\sum_{i\in P, i\notin P'}\mu_i + (\,|\,P\,| - |\,P'\,|\,)c$$

Because $\underline{\nu}(P') \le \underline{\nu}(P)$, the first term is positive. Due to (12), the sum of the second and the third terms is also positive. Therefore, $J^*(\sigma_{P'}) - J^*(\sigma_P) > 0$.□

Proposition 2 indicates that the strategy $\sigma_P$ is guaranteed to be better (i.e. it induces a lower security cost) than another strategy that recommends only a subset $P'$ of $P$ with a lower critical traffic to deploy OTC. It also implies that the critical traffic $\underline{\nu}(P)$ of a subset $P$ of the AS collection is crucial for the performance of the strategy $\sigma_P$ that is associated with this subset.

Using Proposition 2, we are able to determine a structural property for the FDS to be optimal. We call this property the "Maximal Critical Traffic (MCT)" property and define it as follows.

**Definition 2**: An AS collection $\mathcal{N}$ is said to have a "*Maximal Critical Traffic (MCT)*" property if for any $P \subset \mathcal{N}$, the critical traffic of $P$ is no more than that of $\mathcal{N}$, i.e. $\underline{\nu}(P) \le \underline{\nu}(\mathcal{N})$.

We give two examples to illustrate the "MCT" property. Consider a square topology with 4 ASs and symmetric traffic rates between any two connected ASs (Fig. 5(a)). The traffic rates (normalized to GBps) are shown on the edges in the figure.

**Example 1**: $a = 2, b = 3$. The critical traffic of the whole collection is $2a = 4$ (i.e. for AS 1) while all subsets of the AS collection have lower critical traffic. For example, the critical traffic of {2,3,4} is $b = 3$. Therefore, the AS collection has the "MCT" property.

**Example 2**: $a = 1, b = 3$. The critical traffic of the whole collection is $2a = 2$ (i.e. for AS 1) while the critical traffic of the collection $\{2,3,4\}$ is $b = 3$. Therefore, the AS collection does not have the "MCT" property.

The following theorem shows the optimality of the FDS with respect to the "MCT" property.

**Theorem 4**: Under Assumption 3, if the AS collection has the "MCT" property, then the full deployment strategy $\sigma_{\mathcal{N}}$ is optimal.

**Proof**: This is a direct result of Proposition 2 since for any partial deployment strategy, the security cost is increased. □

The "MCT" property implies that the RA has to choose a stronger punishment such that the rating system is IC for any PDS than the FDS. By Proposition 2, this ensures that the PDS induces a higher overall security cost. In Section IV (D), we will study this "MCT" property in more detail by investigating various network topologies.

*C. Iterative Deletion algorithm*

If an AS collection does not have the "MCT" property, then the FDS is not guaranteed to be optimal. Therefore, the RA has to solve for all possible PDSs and as we pointed out, this is a challenging task because of the huge strategy space. However, we may explore the "MCT" property to design a more efficient algorithm to find the optimal strategy, which we present in Table 2. We call this algorithm the *Iterative Deletion (ID)* algorithm.

**Theorem 5**: Under Assumption 3, for an AS collection with $N$ ASs, the Iterative Deletion algorithm solves the optimal rating system design problem within $N$ iterations.

**Proof**: We first consider that in every iteration there is only one AS that has the critical traffic. The analysis can be easily extended to the cases where there are multiple ASs that have the critical traffic. Therefore, in every iteration there is only one AS that is deleted from the subset of ASs that are recommended to deploy OTC. Without loss of generality, we assume that the order of deletion is AS $1,2,...N$ and let $P_i = \{i, i+1, ..., N\}$. Consider any partition $P'$ that does not emerge on the path of the algorithm. Let $j$ be the size of $P'$. Therefore, the smallest AS index $i$ of $P'$ is smaller than $N-j$. Because both $P_i$ and $P'$ contains AS $i$, $\underline{v}(P_i) \geq \underline{v}(P')$. Since it is also true that $P' \subset P_i$, by Proposition 2, $J^*(P_i) < J^*(P')$. □

The basic idea of the ID algorithm is as follows. A milder punishment can be used only when the critical traffic is larger for a subset of the AS collection. Therefore, in each iteration, the ASs with the lowest inbound traffic originating from the remaining AS collection are removed in the hope that the critical traffic of the new collection becomes larger. However, to show that this process does not miss the optimal partition of the AS collection, we also prove that all the other PDSs that do not emerge on the path of the iterative algorithm must incur a higher overall security cost than at least one of the PDSs that is on the path of the iterative algorithm. In Section V (D), we will illustrate how the ID algorithm works for an illustrative example.

*D. Impact of the topology*

In this subsection, we assume that the traffic arrival rates are identical and symmetric on each link between any two connected ASs (denoted by $\lambda_0$). Hence, we focus on the impact of the topology on the optimal rating system design. In this case, the "MCT" property reduces to a "Maximal Critical Degree" property. The following proposition shows that the optimal strategy for many common topologies (please refer to [18] for the exact definitions) is the FDS.

**Proposition 3**: The FDS is optimal if the topology of the AS collection satisfies one of the following conditions: (1) The topology has the identical degree for all nodes, e.g. a "ring lattice" or a complete graph; (2) The topology is a "line", a "star" or a "tree".

**Proof**: It is a direct result of Theorem 4 because these topologies exhibit the "MCT" property and also Assumption 3 automatically holds for an IC rating system when the traffic arrival rates are identical and symmetric on each link. □

In the first case, it is obvious that any subset of the AS collection will have lower critical traffic. In the second case, the connectivity dependency also ensures the "MCT" property. Let us consider the "line" topology as an example, where the critical traffic is injected at the line ends. However, removing the AS at the line end introduces a new line end and hence, the critical traffic is not increased. However, pure network topologies that do not possess the "MCT" property do exist. Among the common topologies discussed in [18], the "core-periphery" may not have the "MCT" property. Therefore, in the following, we focus on the "core-periphery" topology which is widely studied in the context of social and organizational networks [19]. The "core-periphery" topology also deserves special attentions because it is common on the Internet, since it represents the core infrastructure and the local networks.

**Definition 3**: In a core-periphery topology, nodes are divided into two sets: a core set and a periphery set. The connectivity satisfies: (1) Every node in the periphery set connects to some nodes in the core set; (2) Every node in the core set connects to some other nodes in the core set; (3) There is no edge between any two nodes in the periphery set.

Fig. 5(b) shows a "core-periphery" AS collection with 8 ASs. ASs 1-4 are the core ASs and ASs 5-8 are the periphery ASs. We impose some additional restrictions on the "core-periphery" topology in order to study the impact of the AS collection size on the optimal recommended strategy. In the restricted "core-periphery" topology that we consider, there are $N = (1 + l)K$, $(K > 2, l < K)$ ASs in a collection. Among them, $K$ core ASs completely connect to each other. Each of the remaining $lK$ periphery ASs connects to exactly one of the core ASs. The AS collection in Fig. 5(b) also satisfies this additional restriction with $K = 4$ and $l = 1$.

**Proposition 4**: Given the environment parameters $\beta, c, \lambda_0$, there exists a threshold size $N^*(\beta, c, \lambda_0) \geq 0$, such that

1. If $N < N^*$, the optimal strategy is the FDS.
2. If $N \geq N^*$, the optimal strategy is a PDS that recommends only the core ASs to deploy OTC.

**Proof**: The minimal security cost for the FDS is

$$J(\sigma_{full}) = \left[ \underline{p} + g^*(T) \frac{c}{\lambda_0} \right] (K + 2l - 1) JK\lambda_0 + (1 + l)Kc$$

The minimal security cost for the PDS is

$$J(\sigma_{core}) = \left[ \underline{p} + g^*(T) \frac{c}{(K-1)\lambda_0} \right] (K + l - 1) K\lambda_0 + \overline{p} lK\lambda_0 + Kc$$

Their difference can be computed as

$$J(\sigma_{full}) - J(\sigma_{core}) = K \left[ g(T^*)c \left( K + 2l - \frac{l}{K-1} - 2 \right) - \left( (\overline{p} - \underline{p}) l\lambda_0 - lc \right) \right]$$

Note that $K + 2l - l/(K-1) - 2$ increases with $K$ and $(\overline{p} - \underline{p}) l\lambda_0 - lc$ is constant. If $(\overline{p} - \underline{p}) l\lambda_0 - lc > 0$, then there exists a threshold $K^*$, such that for $K \geq K^*$, $J(\sigma_{full}) > J(\sigma_{core})$; for $K < K^*$, $J(\sigma_{full}) \leq J(\sigma_{core})$. If $(\overline{p} - \underline{p}) l\lambda_0 - lc \leq 0$, let $K^* = 0$, the statement also holds. □

The intuition behind Proposition 4 is that when the size is large, a much milder punishment could be used if only the core ASs are recommended to deploy OTC and the increased recovery cost for the traffic originating from the periphery ASs becomes relatively small. Therefore, partial deployment leads to a lower overall security cost. For general topologies that may not have the "MCT" property, the collection size may also play a critical role in determining the optimal recommended strategy. One could imagine that if increasing the AS collection size increases the critical traffic of a PDS, then similar threshold properties may also hold.

## V. Illustrative Results

In this section, we provide numerical results to illustrate the features of our proposed rating system to promote the mutual security of a collection of long-lived ASs. We fix $c = 0.3$, $\overline{p} = 0.3$ and $\underline{p} = 0.05$ and assume that the monitoring error function is $\epsilon(T)=w_0 / (T + 2w_0)$.

*A. Experiments on optimal design parameters $T, p_0$ and the impact of network connectivity*

In this set of simulations, we consider an AS collection with identical connectivity degrees of each AS and identical traffic rates $\lambda_0$ on each edge. However, we vary the degrees $d$ to investigate the impact of the connectivity on the optimal design parameters. Fig. 6 shows that the optimal rating update period emerges as a tradeoff between the monitoring errors and ASs' valuation of the future utilities. When $T$ is small, the monitoring errors are large and hence, the efficiency loss factor is large. When $T$ is large, the punishment is not strong enough to provide incentives for ASs to follow the recommended strategies. Therefore, the optimal $T^*$ is neither too small nor too large. Fig. 7 illustrates the optimal punishment (i.e. filtering qualities for low ratings). When the AS collection becomes denser (i.e. the network degree is larger), ASs obtain more benefits from the OTC technology deployed by others and therefore, their IC constraints are easier to be satisfied. Therefore, a lower $p_0$ can be used. This further leads to a lower overall security cost (normalized to the "first-best" performance) which is illustrated in Table 2. When the monitoring technology is more accurate, the ASs can also receive more benefits if they comply with the recommended strategy and therefore, a lower $p_0$ can also be used. To better illustrate how the optimal design scales with the network connectivity, Fig. 8 shows the optimal filtering quality $p_0$ with respect to various network degrees from 5 to 50.

*B. Performance comparisons*

Fig. 9 illustrates the performance of our proposed optimal rating system design (the solid curve with square markers) with the benchmarks introduced in Section II (C). The simulated AS collection consists 8 fully connected ASs with identical traffic $\lambda_0 = 1$ between any two of them. The discount parameter $\beta = 0.2$ is utilized in this simulation. In this case, the FDS is the optimal strategy. We vary the monitoring error (in terms of $w_0$) to investigate the impact of the imperfect monitoring on the achievable security level. The optimal design significantly outperforms the three benchmarks ("no OTC", "rating-independent OTC", "worst-best") and also a rating system with a fixed design. Note that "rating-independent OTC" (i.e. the filtering qualities do not depend on ratings) achieves as high as a security cost by "no OTC" (i.e. no OTC is deployed) since no ASs have the incentives to deploy OTC. “Worst-best” ( $\underline{p}$ for high ratings and $\overline{p}$ for low ratings) is able to provide ASs with incentives to deploy OTC. However it induces a significant efficiency loss due to the imperfect monitoring. When the monitoring error is large ( $w_0 > 0.35$ ), the fixed design cannot provide sufficient incentives for the ASs to make investment and hence, the overall security cost is the highest. When the monitoring error is small ( $w_0 < 0.35$ ), the fixed design can provide sufficient incentives but is still much worse than the optimal design.

*C. Experiments on rating systems, Tit-for-Tat and trigger strategies*

If two ASs have mutual traffic between each other, direct reciprocation schemes such as a Tit-for-Tat (TfT) scheme may also provide ASs with incentives to make security investments. In a TfT scheme, if an AS $i$ sends unfiltered traffic to another AS $j$, then in the next period AS $j$ also sends unfiltered traffic to AS $i$ as a punishment. However this direct reciprocity may not exist if there is only traffic from AS $i$ to AS $j$ and hence, a TfT scheme does not work. Even if we assume direct reciprocity exists for every pair of ASs, a TfT scheme still has a significant efficiency loss as shown in Fig. 10. The TfT scheme generates the similar performance as the proposed rating system when the discount factor parameter is small (ASs value more the future utilities) but does not work when the discount factor parameter is large (ASs value less the future utilities). However, the proposed rating system is able to achieve a low overall security cost even when ASs value less the future utilities. When the monitoring is perfect, trigger strategies are the strongest punishment and provide the strongest incentives for ASs to deploy OTC. However, in the imperfect monitoring scenarios as in Fig. 10, the AS collection will eventually be in a punishment phase where all ASs do not deploy OTC anymore in the future. Therefore, the performance of the trigger strategies significantly degrades in the imperfect monitoring scenarios.

*D. Illustration of the ID algorithm*

In this experiment, we show how the ID algorithm works for a specific deployment scenario. Consider the AS collection shown in Fig. 11. For simplicity, we assume symmetric traffic arrival rates between every two ASs. The traffic rates (normalized to GBps) are shown on the edges in the figure. Table 3 illustrates the 5 iterations of the ID algorithms. The first iteration evaluates the performance of the FDS. The critical traffic is 2 which is the inbound traffic of AS1. In the second iteration, AS 1 is deleted from the set of ASs that are recommended to deploy OCT.

Because the traffic from AS 1 to AS 2 is unfiltered, the critical traffic becomes 3 which is the AS 2's aggregate inbound traffic originating from the remaining ASs. In the third iteration, AS 2 is deleted and AS 3 has the critical traffic. In the fourth iteration, deleting AS 3 does not increase the critical traffic and hence, the overall security cost must be larger. AS 5 has the critical traffic. In the last iteration, deleting AS 5 decreases the critical traffic and hence, the overall must be even larger. The algorithm stops at the fifth iteration because there are only two ASs left. Therefore, the minimal security cost is achieved when ASs 3,4,5,6 are recommended to deploy OTC while ASs 1, 2 are recommended not to. The minimal overall security cost is 6.74.

*E. Impact of collection size for "core-periphery" topologies*

We consider the restricted "core-periphery" AS collections with identical traffic arrival rates and $l = 1$ but vary the size of the AS collection. Fig. 12 shows the minimal security costs achieved by the FDS and the PDS that recommends only the core ASs deploy OTC. As the AS collection size becomes larger, the FDS induces a higher normalized security cost while the PDS induces a lower normalized security cost. In this example, the threshold is $N^* = 22$. If the AS collection size is smaller than 22, then the FDS is optimal. If the AS collection size is larger than 22, then the PDS is optimal.

## VI. Related Works

Network security has become one of the major focuses of network science [1][2][3]. Interconnected and interdependent networks often share security risks and therefore, establishing a secure network environment requires security investment and security technology deployment from all interconnected networks. A major source of complication in network security is the autonomous and self-interested nature of decision making of the networks. This gives rise to a non-cooperative game when the networks decide their security investment policies.

The game theoretical framework is becoming an important tool for network modeling and design and has been employed in various networking contexts [15][13][12]. Our paper builds on existing research studying the security investment of a collection of interdependent ASs [6]. This literature generally can be classified into two categories. The first category only characterizes the performance loss of the interconnected ASs at equilibrium, but does not design incentive schemes to achieve the socially optimal security level. Among these works, the positive externality effect of the security investment is studied in [5][6]. Another strand of literature considers inoculation strategies in epidemic networks [11][12][13]. The ASs' interactions are often modeled as a one-shot game for which the Nash equilibrium (NE) is determined and the Price-of-Anarchy (i.e. the ratio of the optimum social welfare to the worst NE social welfare) is computed. Hence, these works do no design incentives to enhance the performance of the NE.

The second category designs incentive schemes to encourage security investment. One solution to encourage security investments is to adopt cyberinsurance [14], which is proposed to transfer residual security risks to an insurance company. However, such solutions are complex (e.g. insurance premium with discriminations) and moreover, are not able to achieve the social optimum. Certification [7] is another way to provide security investment incentives. However, it requires monetary transfers between ASs which might be difficult to perform in practice and also induce strict social efficiency losses since certification subscription fees are required. A limitation of this strand of literature is that the interaction among ASs is modeled as one-shot games, thereby disregarding the repeated nature of their interaction.

Rating or reputation schemes are widely applied to deal with incentive problems in online communities with self-interested users which are interacting repeatedly (i.e. are long-lived) [15][16][17]. Nevertheless, none of these schemes can be directly applied to cybersecurity problems because they fail to incorporate many specific features that are critical to characterize the ASs and the cyber-environment, which we highlight in Table 4.

## VII. CONCLUSIONS

In this paper, we studied the design problem of rating systems aimed at encouraging security investment within a collection of ASs that interact regularly and over a long period of time. We showed that it is possible to exploit the ongoing nature of the interaction to design rating systems to improve the mutual security. Our analysis showed that the traffic and the connectivity structure of the AS collection can strongly influence ASs self-interested investment decisions. Surprisingly, a lower security cost may be achieved by recommending partial deployment of the security technology than by recommending full deployment. However, whether this improvement can be achieved depends on the structure of the AS collection and the monitoring capability of ASs' investment actions. As a next step, the proposed rating systems can be used to design security policies that can deal with a variety of other security problems (risk management, risk-driven routing policies, deployment of firewalls etc.) besides the one considered in this paper.

## REFERENCES

[1] R. Anderson, “The economics of information security,” Science, 317, 610 (2006).

[2] C. Partridge, “Forty data communications research questions,” ACM SIGCOMM Computer Communication Review, Vol. 41, Sept. 2011.

[3] D. Talbot, “The Internet is broken,” Technology Rev., Dec. 2005/Jan. 2006.

[4] H. Varian, “System reliability and free riding,” Economics of Information Security, Springer, 2004.

[5] M. LeLarge, “Economics of malware: epidemic risks model, network externalities and incentives,” Allerton 2009.

[6] H.Kunreuther and G. Heal, “Interdependent security,” Journal of Risk and Uncertainty, vol. 26, no. 2-3, 2003.

[7] M. Parameswarn, X. Zhao, A. Whinston, and F. Fang, “Reengineering the Internet for better security,” Computer, 40(1), pp. 40-44, Jan. 2007.

[8] S. Shetty, “Protocol-level malware scanner,” US Patent, 2004.

[9] Y. Takahashi, K. Ishibashi, “Incentive mechanism for prompting ISPs to implement outbound filtering of unwanted traffic,” Intl. Conf. Network games, control and optimization (NetGCOOP), 2011.

[10] M. Bloem, T. Aplean, S. Schmidt, “Malware filtering for network security using weighted optimality measures,” IEEE International Conference on Control Applications (CCA), 2007.

[11] J. Aspnes, K. L. Chang, and A. Yampolskiy, “Inoculation strategies for victims of viruses and the sum-of-squares partition problem,” J. Comput. Syst. Sci. 72(6): 1077-1093, 2006.

[12] V. S. A. Kumar, R. Rajaraman, Z. Sun, R. Sundaram, “Existence theorems and approximation algorithms for generalized network security games,” IEEE ICDCS, 2010

[13] J. Omic, A. Orda, P. van Mieghem, “Protecting against network infections: a game theoretic perspective,” IEEE INFOCOM 2009.

[14] J. Bolot, M. Lelarge, “A new perspective on Internet security using Insurance,” IEEE INFOCOM, 2008.

[15] F. Millan, J. Jaramillo, R. Srikant, “Achieving cooperation in multihop wireless networks of selfish nodes,” Gamenets 2006.

[16] C. Dellarocas, “Reputation mechanisms design in online trading environments with pure moral hazard,” Information Systems Research, Vol. 16, No. 2, pp. 209-230, June 2005.

[17] Y. Zhang and M. van der Schaar, “Reputation-based incentive protocols in crowdsourcing applications,” IEEE INFOCOM 2012.

[18] E. Airoldi, “Sampling algorithms of pure network topologies: stability and separability of metric embeddings,” ACM SIGKDD Explorations Newsletter, vol. 7, issue. 2, Dec 2005.

[19] S. Borgatti, M. Everett, “Models of core/periphery structures,” Social Networks 21(1999) 375-395, Elsevier.

[20] S. Nageeb Ali, D. Miller, “Enforcing cooperation in networked societies,” working paper, 2010.

[21] G. Mailath and L. Samuelson, Repeated games and reputations: long-run relationships, Oxford University Press, 2006.

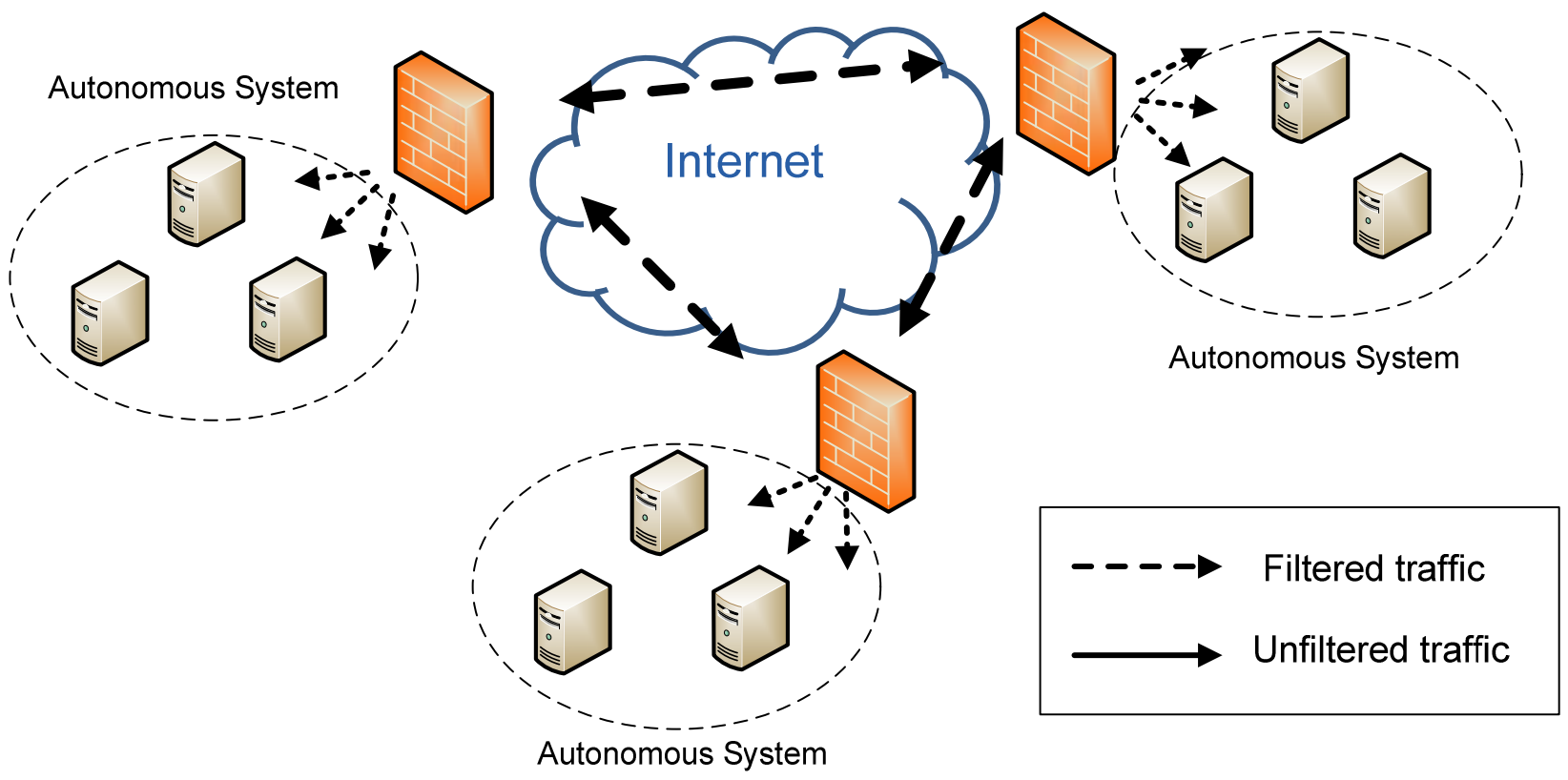

Fig. 1 An illustrative system with outbound traffic control. Each AS not only filters inbound traffic from the Internet but also filters outbound traffic to the Internet

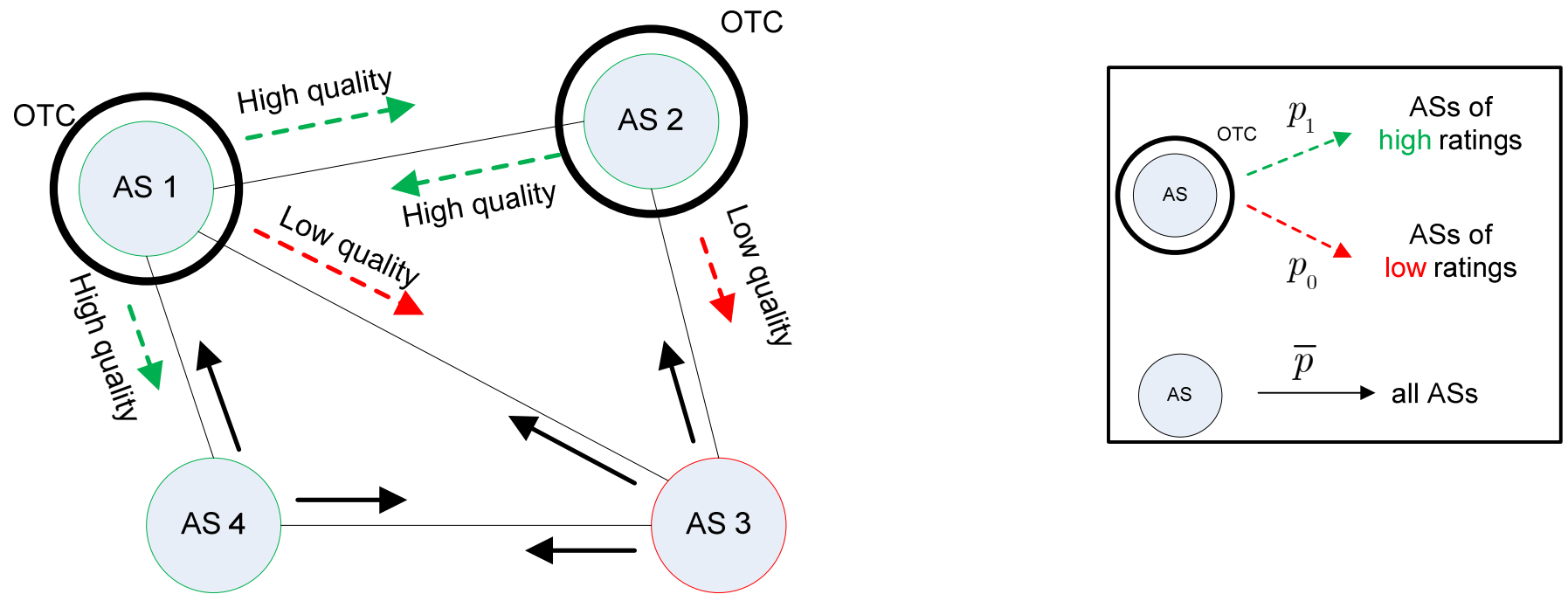

Fig. 2 Illustration of ASs' traffic exchange and filtering interactions.
(AS 1,2,3 are recommended to deploy OTC while AS 4 is recommended to not deploy OTC. The only AS that deviates is AS 3.)

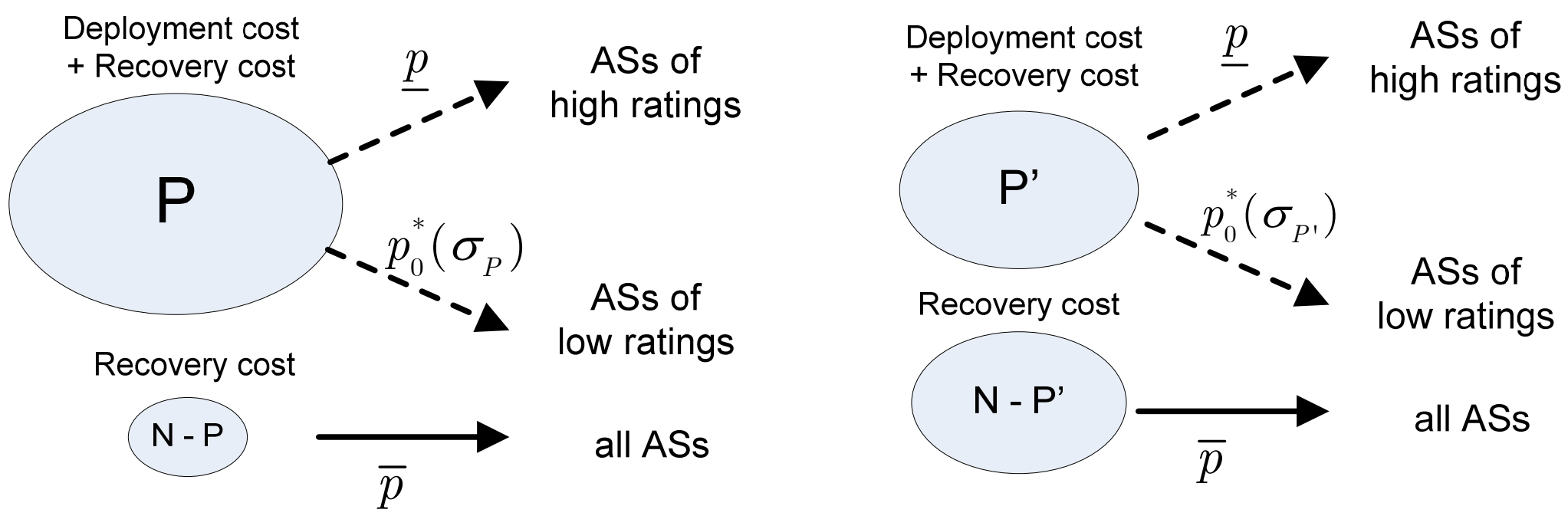

Fig. 3 Overall security costs for given recommended strategies.
(For $\sigma_{P'}$, the recovery cost for the traffic originating from ASs that do not deploy OTC increases. The deployment cost for the ASs that deploy OTC decreases. Whether or not the recovery cost for the traffic originating from ASs that deploy OTC decreases depends on the critical traffic. )

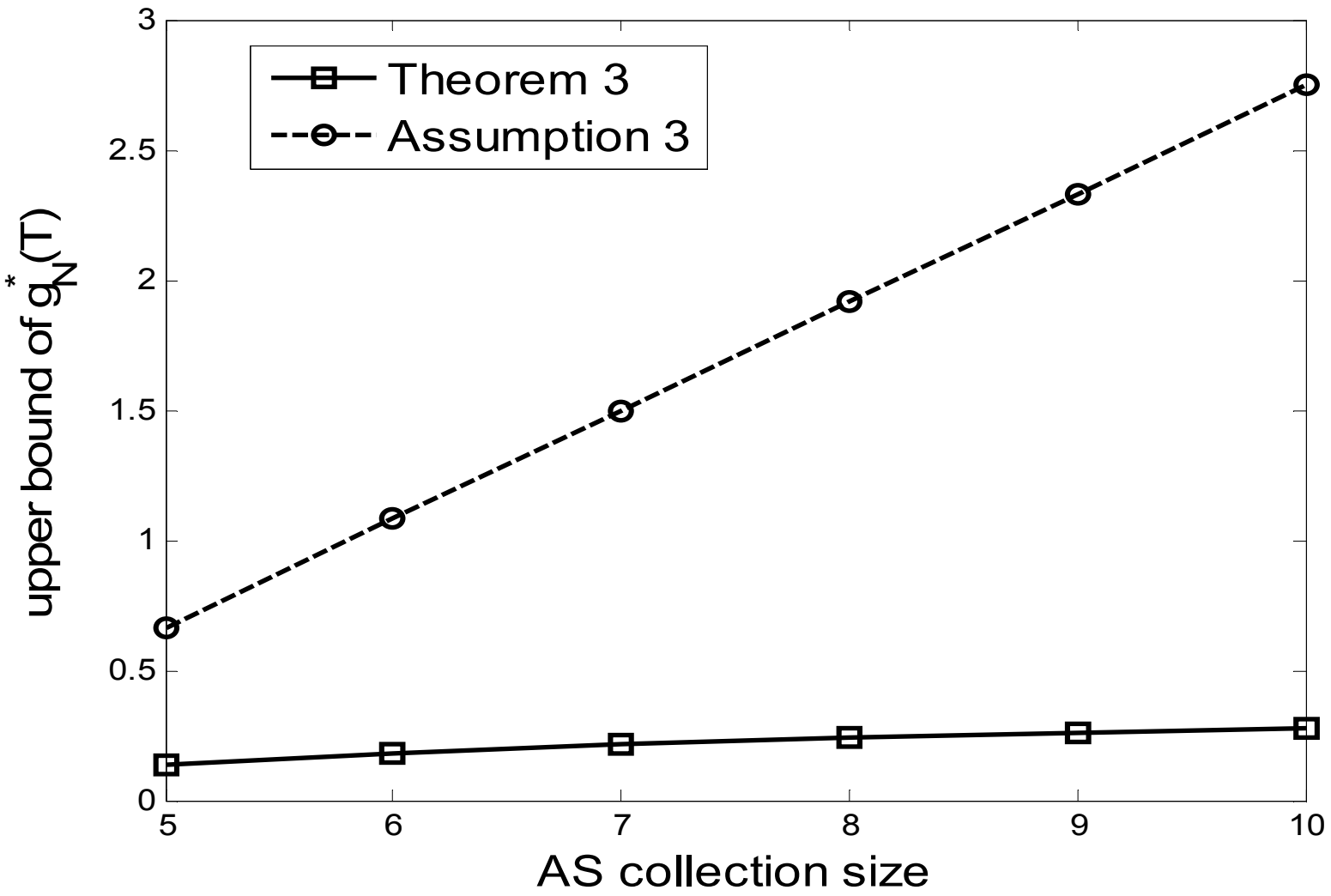

Fig. 4 Illustration of the bound in Assumption 3.

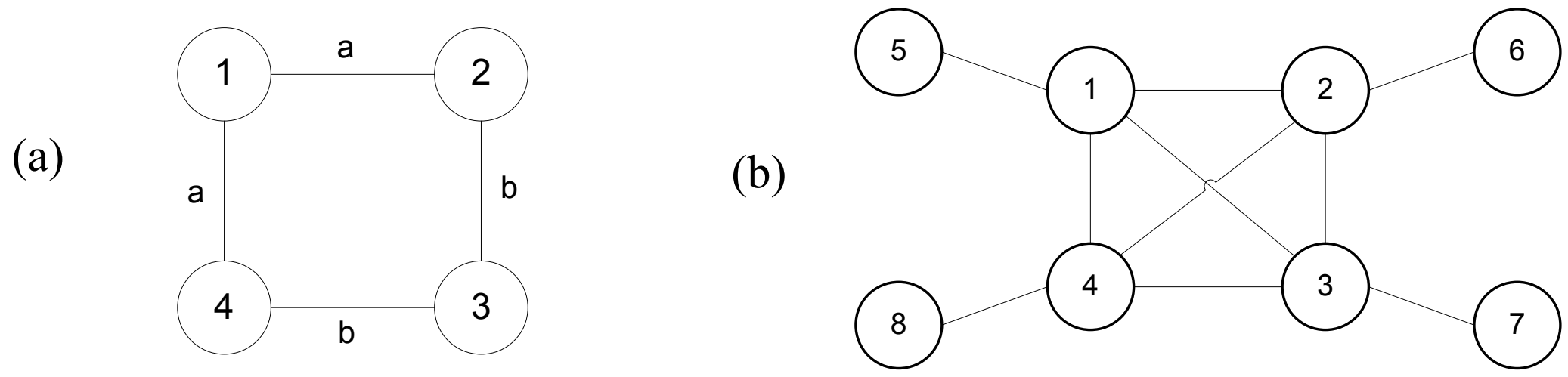

Fig. 5 (a) Illustration of the "Maximal Critical Traffic" property.
(b) An illustrative "core-periphery" topology.

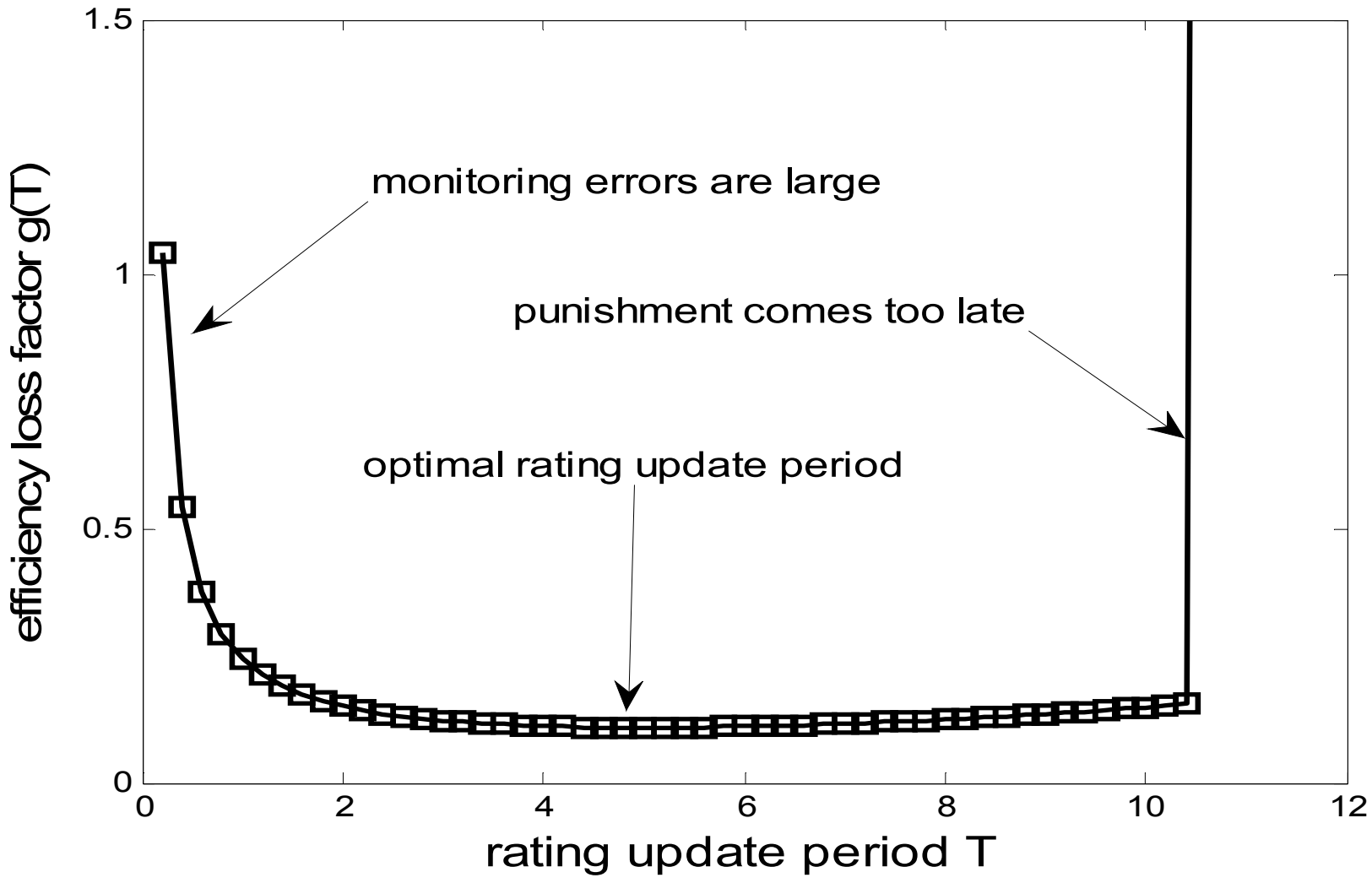

Fig. 6 Optimal rating update period. ($\overline{p} = 0.3, \underline{p} = 0.05, c = 0.3, \beta = 0.4, \lambda_0 = 1, w_0 = 0.2$).

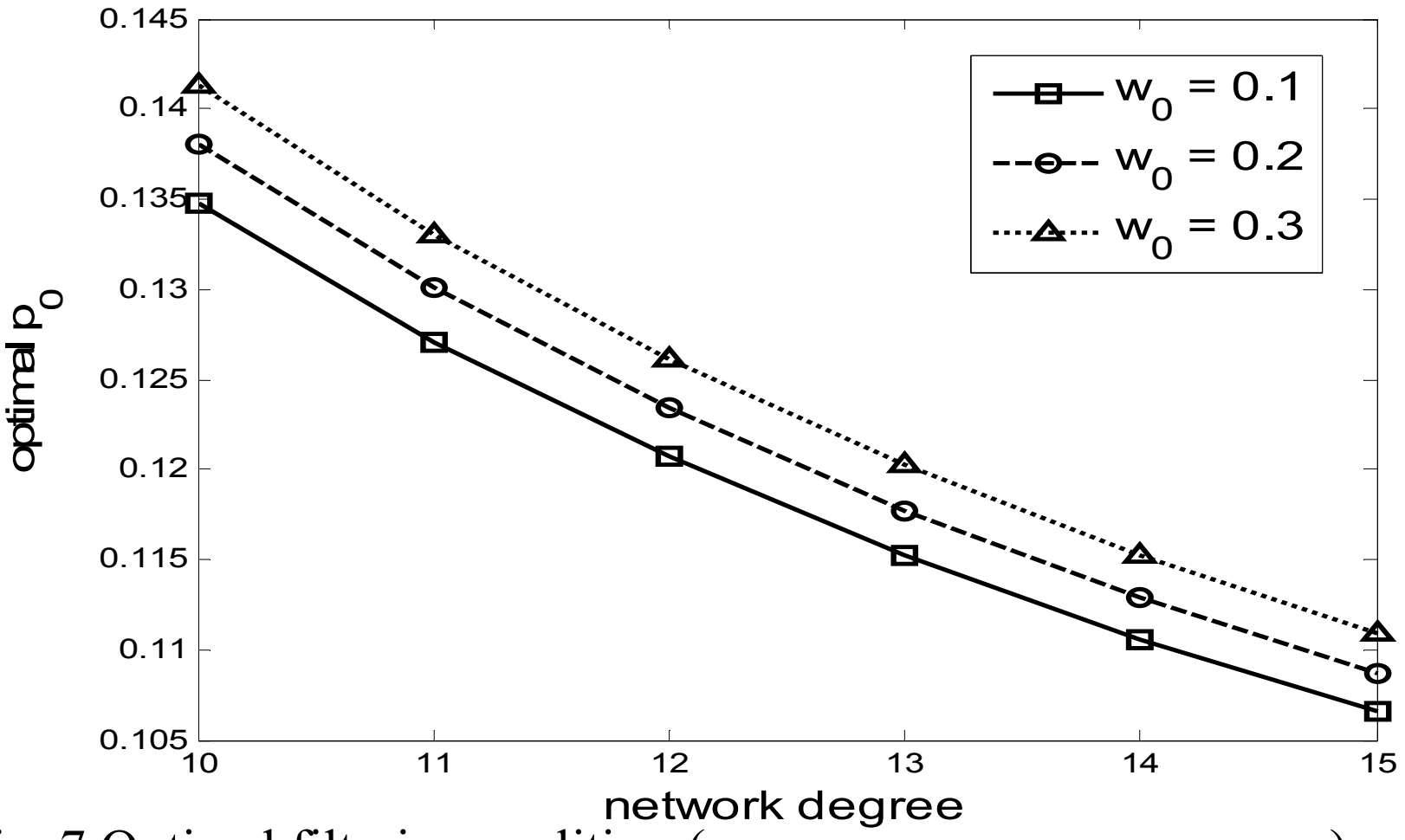

Fig. 7 Optimal filtering qualities. ($\overline{p} = 0.3, \underline{p} = 0.05, c = 0.3, \beta = 0.4, \lambda_0 = 1$).

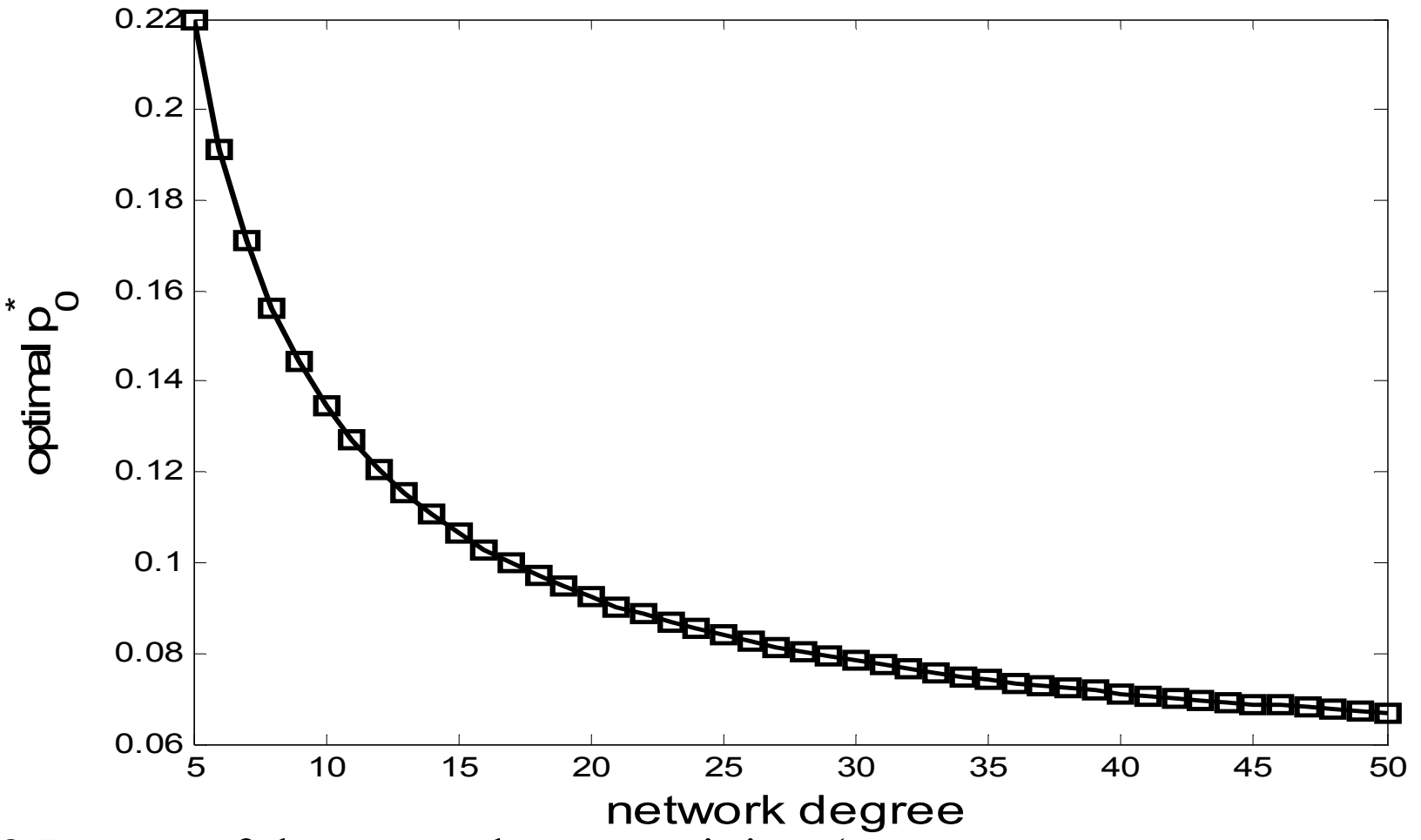

Fig. 8 Impact of the network connectivity. ($\overline{p} = 0.3, \underline{p} = 0.05, c = 0.3, \beta = 0.4, \lambda_0 = 1, w_0 = 0.1$).

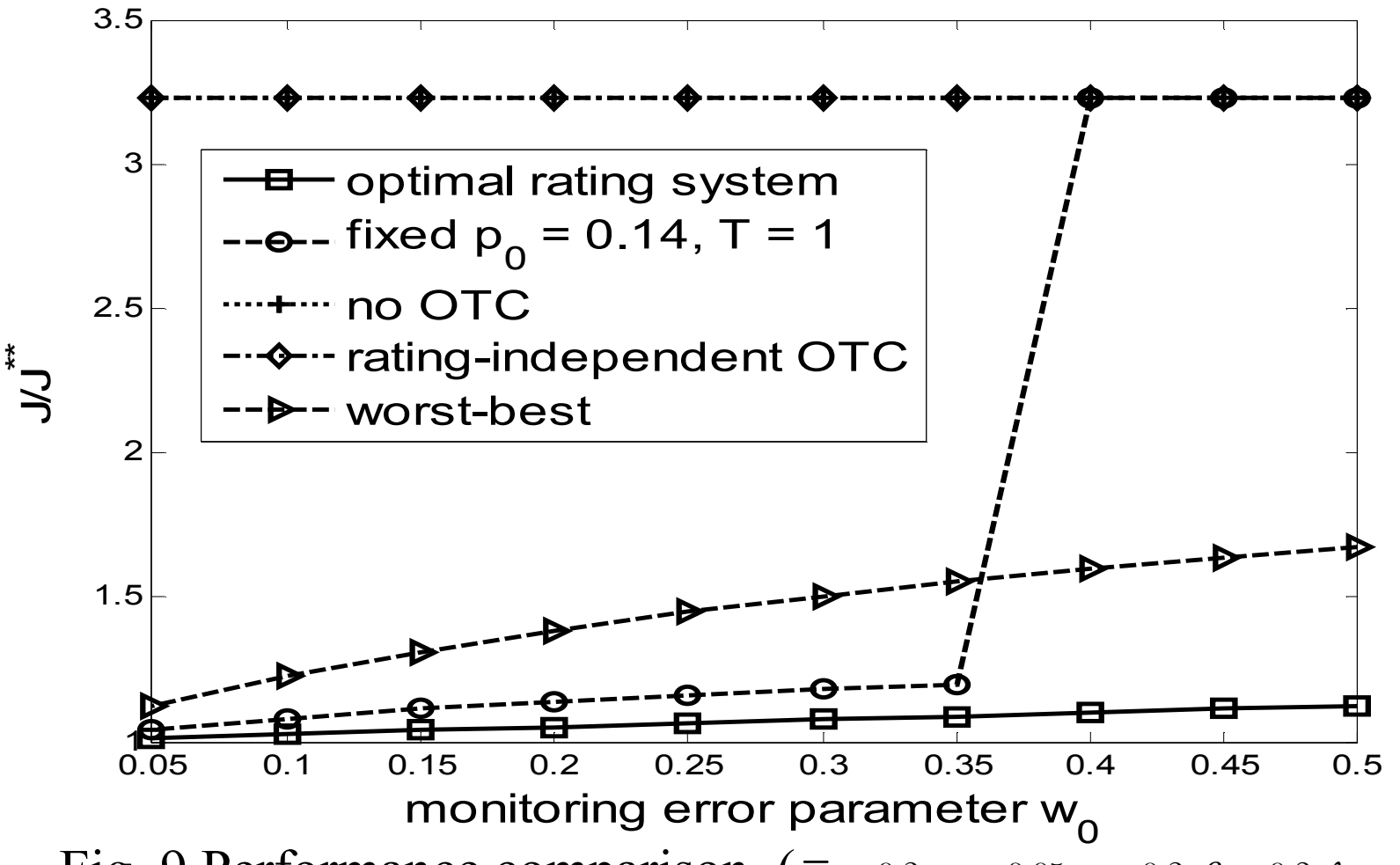

Fig. 9 Performance comparison. ($\overline{p} = 0.3, \underline{p} = 0.05, c = 0.3, \beta = 0.2, \lambda_0 = 1, N = 8$).

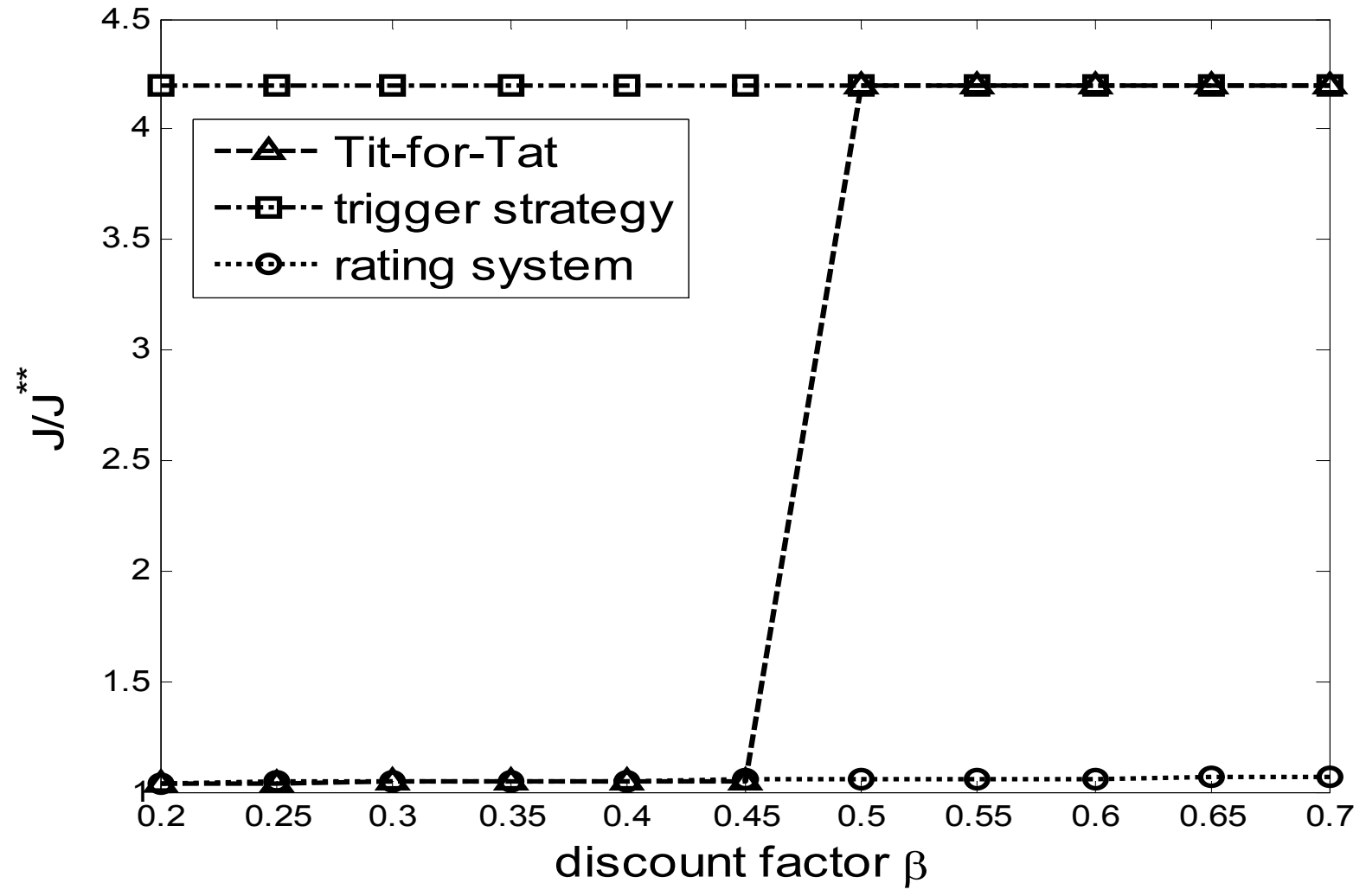

Fig. 10 TfT, trigger strategies vs. rating systems. ($\overline{p} = 0.3, \underline{p} = 0.05, c = 0.3, \lambda_0 = 2, N = 8, T = 1, w_0 = 0.1$)

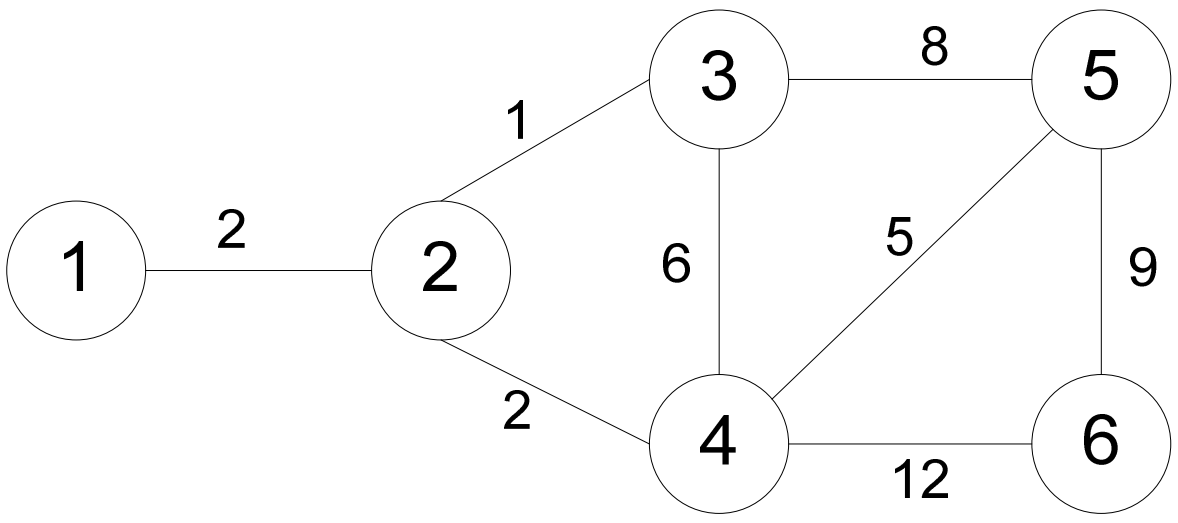

Fig. 11 An illustrative AS collection.

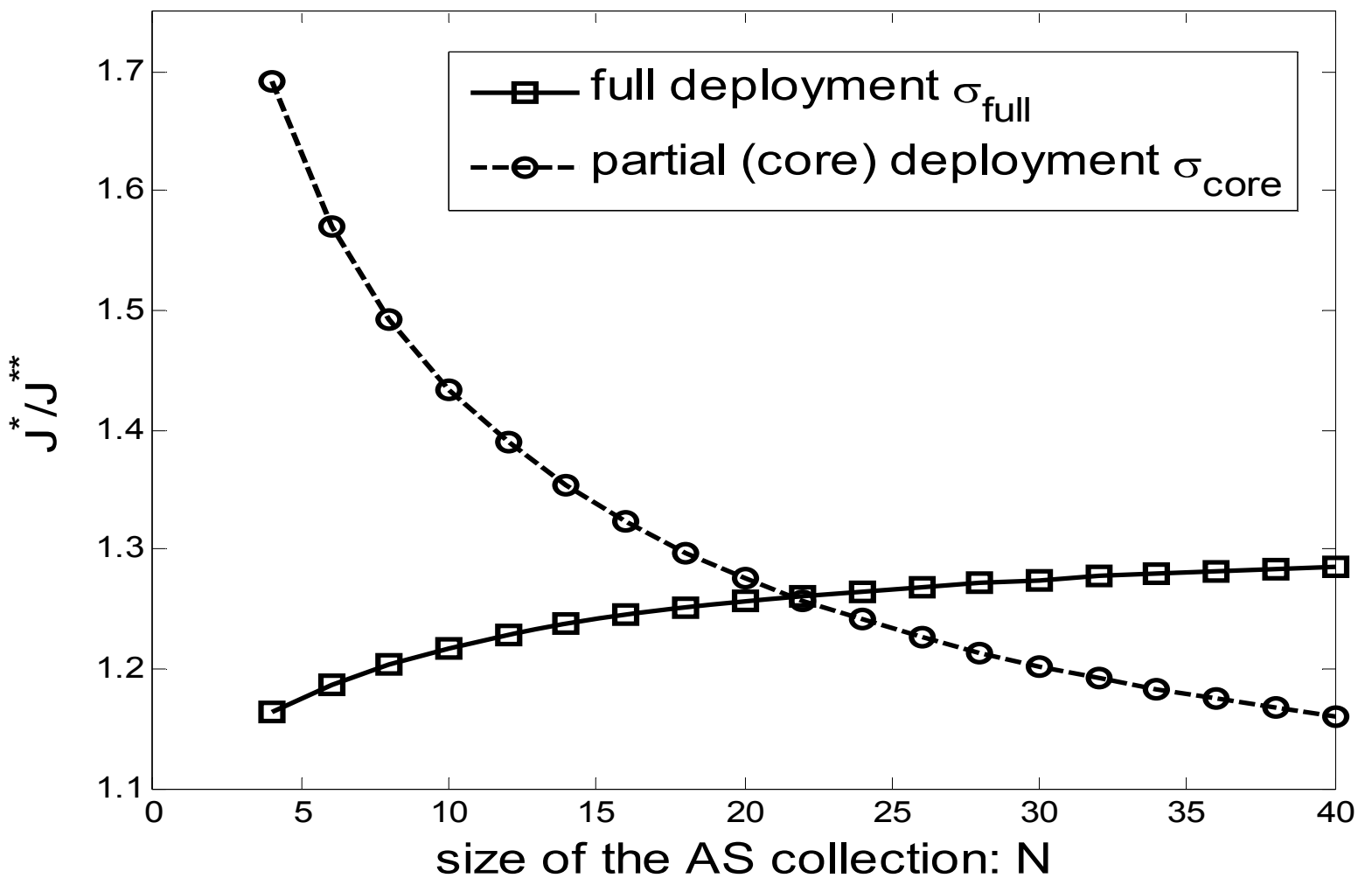

Fig. 12 Critical AS collection size. ($\overline{p} = 0.3, \underline{p} = 0.05, c = 0.3, \beta = 0.2, \lambda_0 = 4, w_0 = 0.4$)

*Table 1: Iterative Deletion (ID) Algorithm*

**Input**: AS collection $\mathcal{N}$, traffic matrix $\Lambda$
**Output**: Optimal rating system design, $p_0^*, T^*, \sigma^*$

Set $T^* = \arg\min_T g(T)$;

Solve the optimal rating system design for $\sigma_{\mathcal{N}}$;

Set $p_0(0) = p_0^*(\sigma_{\mathcal{N}})$, $J(0) = J^*(\sigma_{\mathcal{N}})$, $\sigma(0) = \sigma_{\mathcal{N}}$;

Set $P = \mathcal{N}$, iteration index $t = 1$;

while $P \neq \varnothing$, do:

    Let $\tilde{P} = P - \{i : v_i(P) = \underline{v}(P)\}$;

    If $\underline{v}(P) \leq \underline{v}(\mathcal{N})$

        Set $P = \tilde{P}$; continue;

    end if

    If $\underline{v}(P) > \underline{v}(\mathcal{N})$

        Set $\sigma(t) = \sigma_P$;

        Solve the optimal rating system design for $\sigma_P$;

        Set $p_0(t) = p_0^*(\sigma_P)$, $J(t) = J^*(\sigma_P)$;

        $t = t + 1$;

    end if

end while

Let $t^* = \arg\min_t J(t)$;

$\sigma^* = \sigma(t^*), p_0^* = p(t^*)$

| | **d = 5** | **d = 6** | **d = 7** | **d = 8** | **d = 9** | **d = 10** |
|---|---|---|---|---|---|---|
| **$w_0$ = 0.1** | 1.0204 | 1.0192 | 1.0181 | 1.0172 | 1.0163 | 1.0155 |
| **$w_0$ = 0.2** | 1.0408 | 1.0384 | 1.0362 | 1.0343 | 1.0326 | 1.0311 |
| **$w_0$ = 0.3** | 1.0612 | 1.0576 | 1.0544 | 1.0515 | 1.0489 | 1.0466 |

Table 2. The impact of AS collection density on the overall security costs (normalized to the "first-best" performance).

| t | AS 1 | AS 2 | AS 3 | AS 4 | AS 5 | AS 6 | J |
|---|---|---|---|---|---|---|---|
| 1 | ***2*** | *5* | *15* | *25* | *22* | *21* | 7.58 |
| 2 | -- | ***3*** | *15* | *25* | *22* | *21* | 6.98 |
| 3 | -- | -- | ***14*** | *23* | *22* | *21* | 6.74 |
| 4 | -- | -- | -- | *17* | ***14*** | *21* | >6.74 |
| 5 | -- | -- | -- | *12* | -- | *12* | >6.74 |

Table 3. Iterations of the ID algorithm.

| | [15] | [16] | [17] | Proposed |
|---|---|---|---|---|
| *Players interaction* | one-to-one (fixed) | one-to-one (changing) | one-to-one (rnd. match) | many-to-many |
| *Reciprocation type* | direct | indirect | indirect | indirect/direct |
| *Rating update based on* | single event | single/a series of event(s) | single event | a series of events |
| *Cost/benefit heterogeneity* | NA | NA | no | yes (different traffic rates) |
| *Underlying topology* | no | no | no | yes (connectivity) |
| *Recommended strategy* | homogeneous | homogeneous | homogeneous | player-specific |
| *Imperfect Monitoring* | yes | yes | no | yes |

Table 4. Comparisons of the existing works and the proposed rating systems.